\pdfoutput=1

\def\editmode{0}
\def\reportmode{0}

\def\bibfilenames{Abbrev,WISENET,ChannelGains}

\if\reportmode1
    \documentclass[10pt,final,onecolumn]{IEEEtran}
\else
\documentclass[10pt,conference]{IEEEtran}
    \IEEEoverridecommandlockouts
\fi

\if\editmode1  
\usepackage[backend=bibtex,style=alphabetic,sorting=debug]{biblatex}
\DeclareFieldFormat{labelalpha}{\thefield{entrykey}}
\DeclareFieldFormat{extraalpha}{}
\bibliography{\bibfilenames}

\newcommand{\cmt}[1]{\noindent\textcolor{lightgreen}{\underline{[#1]}}} 
\newenvironment{myitemize}{\begin{itemize}}{\end{itemize}}
\newcommand{\myitem}{\item}

\newcommand{\acom}[1]{\textcolor{red}{[#1]}}
\newcommand{\nextver}[1]{{\color{orange}{#1}}}
\newcommand{\rev}[1]{\textcolor{blue}{#1}}
\newcommand{\notation}[1]{\textcolor{darkgreen}{#1}}
\else
\usepackage{cite}
\newcommand{\cmt}[1]{} 
\newenvironment{myitemize}{}{}
\newcommand{\myitem}{}

\newcommand{\acom}[1]{}
\newcommand{\nextver}[1]{}
\newcommand{\rev}[1]{\textcolor{blue}{#1}}
\renewcommand{\rev}[1]{{#1}}
\newcommand{\notation}[1]{#1}

\fi

\usepackage{algorithm}
\usepackage{stfloats}
\usepackage[noend]{algpseudocode}

\title{Channel Gain Cartography via Mixture of Experts}

\newcommand{\theAck}{The authors want to acknowledge Assoc. Prof. Daniel Romero for his participation in discussions during an early stage of this work.}

\if\reportmode1
    \author{
        Luis Miguel Lopez-Ramos, 
        Yves Teganya, 
        Baltasar Beferull-Lozano, and 
        Seung-Jun Kim
        \thanks{\theAck}
        \\[.5cm]\today
    }
\else
\author{Luis M. Lopez-Ramos,~\IEEEmembership{Member,~IEEE,} Yves Teganya,~\IEEEmembership{Student Member,~IEEE,} \\ Baltasar Beferull-Lozano,~\IEEEmembership{Senior Member,~IEEE,} and Seung-Jun Kim,~\IEEEmembership{Senior Member,~IEEE}
\thanks{This work was supported by grant(s) FRIPRO TOPPFORSK WISECART grant 250910/F20 and SFI Offshore Mechatronics 237896/E30, 
from the Research Council of Norway, as well as by US NSF grant 1547347. 
}
\thanks{The first three authors  are  with  the  WISENET  Center,  Dept.  of  ICT,  University  of Agder,  Jon  Lilletunsvei  3,  Grimstad,  4879  Norway.  E-mails:\{luismiguel.lopez, yves.teganya, baltasar.beferull\}@uia.no. Seung-Jun Kim is with the Dept. of Comput. Sci. Electr. Engr., University of Maryland, Baltimore County, Baltimore, MD 21250, USA. E-mail: sjkim@umbc.edu.}
}
\fi

\usepackage{fixltx2e}

\usepackage{graphicx}

\usepackage{subcaption}              

\usepackage[utf8]{inputenc}

\usepackage{amsfonts}
\usepackage{amsmath}
\usepackage{mathtools}

\usepackage{amssymb}

\usepackage{bm}

\usepackage{color,verbatim}
\usepackage{multirow}
\usepackage{accents}

\usepackage{theoremref}

\usepackage{url}

\newcounter{rulecounter}
\newcommand{\resetrule}{ \setcounter{rulecounter}{0}}
\resetrule

\newsavebox{\selvestebox}
\newenvironment{colbox}[1]
  {\newcommand\colboxcolor{#1}%
   \begin{lrbox}{\selvestebox}%
   \begin{minipage}{\dimexpr\columnwidth-2\fboxsep\relax}}
  {\end{minipage}\end{lrbox}%
   \begin{center}
   \colorbox{\colboxcolor}{\usebox{\selvestebox}}
   \end{center}}

\definecolor{orange}{rgb}{1,0.8,0}
\definecolor{gray}{rgb}{.9,0.9,0.9}
\definecolor{darkgray}{rgb}{.3,0.3,0.3}
\definecolor{darkblue}{rgb}{.1,0.0,0.3}
\definecolor{lightblue}{rgb}{0.7,0.7,1}
\definecolor{lightred}{rgb}{1,0.7,.7}
\definecolor{purple}{RGB}{204,153,255}
\definecolor{lightgray}{rgb}{.95,0.95,0.95}
\definecolor{lightgreen}{rgb}{0.3,0.5,0.3}
\definecolor{darkgreen}{rgb}{0.05,0.3,0.05}

\newcommand{\ra}{$\rightarrow$~}

\newtheorem{myproposition}{Proposition}
\newtheorem{myremark}{Remark}
\newtheorem{myproblemstatement}{Problem Statement}
\newtheorem{mylemma}{Lemma}
\newtheorem{mytheorem}{Theorem}
\newtheorem{mydefinition}{Definition}
\newtheorem{mycorollary}{Corollary}

\begin{document}

\maketitle

\begin{abstract}
In order to estimate the channel gain (CG) between the locations of an arbitrary transceiver pair across a geographic area of interest, CG maps can be constructed from spatially distributed sensor measurements. 
Most approaches to build such spectrum maps are \emph{location-based}, meaning that the input variable to the estimating function is a pair of spatial locations. The performance of such maps depends critically on the ability of the sensors to determine their positions, which may be drastically impaired if the positioning pilot signals are affected by multi-path channels. 
An alternative \emph{location-free} approach was recently proposed for spectrum power maps, where the input variable to the maps consists of features extracted from the positioning signals, instead of location estimates. The location-based and the location-free approaches have complementary merits. In this work, apart from adapting the location-free features for the CG maps, a method that can combine both approaches is proposed in a mixture-of-experts framework.
\end{abstract}

\section{Introduction}

\cmt{Spectrum Cartography: Why is it important}\begin{myitemize}
    \myitem \cmt{D2D comms}
    Information regarding the channel gain (CG) between a pair of wireless transceivers is critical in a plethora of resource allocation (RA) algorithms. In the context of device-to-device (D2D) communication and cognitive radios, judiciously designed RA algorithms can boost the network performance metrics significantly~\cite{elnourani2018underlay}. For instance, consider a cellular system where regular cellular connections coexist with D2D communication. A RA scheme can be implemented for channel assignment and/or power allocation across the pairs of D2D devices, with the goal of maximizing the total aggregated throughput or other relevant metrics. Such a RA scheme will typically require estimates of the CGs between the cellular users and the D2D users, and between the transmitter and the receiver in each of the D2D links, in order to quantify the expected interference caused from sharing the communication channels. Given the difficulty of continuously measuring the CGs between arbitrary pairs of devices, the approach based on the CG map is very useful~\cite{kim2011cooperative,lee2019bayesian}. The RA performance depends heavily on the accuracy of the CG estimates; therefore, the accuracy of the map is critical. Moreover, in the case of mobile networks, the CGs for the spatial locations where the transceivers are expected to be in the future time slots are also necessary. Thus, it is important to have CG estimates not only where the transceivers are currently located, but also in the arbitrary locations around them.
    
    \myitem \cmt{Mapping in the math sense} In its simplest form, a CG map can be defined as a function that maps a pair of spatial locations to an estimate of the CG between them. 
    The CG maps were estimated by recovering the so-called spatial loss field~\cite{agrawal2009correlated}, where the gain was modeled as a distance-based loss plus a weighted integral of the spatial loss field\nextver{More modern approaches such as [check Donghoon Lee's thesis] develop improvements such as...}. This model accounts for the loss due to the absorption from obstacles, but can be inaccurate in the multi-path propagation environments where the signals may get severely attenuated, or amplified by multiple reflections. 
    
The data samples used to train the CG function are taken from a set of pairs of sensing nodes spread across the area to be covered by the map. Thus, the existing methods typically rely on accurate sensor location estimates~\cite{kim2011cooperative,lee2019bayesian}. However, under practical conditions (for instance, when the localization is achieved using positioning pilots sent by base stations), the sensing nodes might not be able to determine their locations accurately. Specifically, the time (difference) of arrival (ToA/TDoA) measurements may experience severe bias due to non-line-of-sight (NLoS) conditions\cite{wang2016robust}, with large localization error as a consequence. The training data is thus noisy in the input variable, which then translates into errors in the map estimates. The error due to the NLoS propagation in strong multi-path environments may even destroy many location estimates, rendering the CG map uninformative.
\end{myitemize}

\cmt{LocBased} 
 
\cmt{LocFree} 
\cmt{First mention of LocFree maps} Recently, in the simpler case of an interference power map, which is a function of a single location, 
this issue was mitigated by extracting features from the pilot signals directly, and building the maps using kernel ridge regression (KRR)~\cite{teganya2019location}. Such features have nature similar (but not exactly equal) to the positioning features such as the ToA or TDoA. Following \cite{teganya2019location}, here the (standard) procedure of estimating the location of the sensors from the available pilot signals, and training the map as a function of a vector of spatial coordinates, is referred to as the ``location-based" (LocB) cartography. On the other hand, the (novel) method of directly learning the map estimate as a function of the pilot signal features is referred to as the ``location-free" (LocF) cartography. 

\cmt{Focus on LocB} While many works explored the use of advanced regression techniques (such as those based on deep learning \cite{han2020power, ye2018channel} or advanced kernel methods \cite{aldossari2019machine}) to improve the accuracy of spectrum maps, most of these works build LocB maps. When the estimation of the sensor locations is accurate, the maps produced by these methods are also accurate. Consequently, these methods rely on pilot signals from base stations only when the LoS component is strong, but not in strong multi-path environments.
 \nextver{Balta: It would be great to show here some spectrum map where one can observe regions where Location-based algorithm works well and other regions where it does not work well}
\nextver{Some recent approaches based on Generative Models (GANs):
- X. Han, L. Xue, F. Shao, and Y. Xu, “A Power Spectrum Maps Estimation Algorithm Based on Generative Adversarial Networks for Underlay Cognitive Radio Networks,” Sensors, vol.20,no.1,p 311,2020. 
- S.M.Aldossari and K.C.Chen,“Machine learning for wireless communication channel modeling:An overview,” Wireless PersonalCommunications, vol.106,no.1,pp.41–70,2019.
- H. Ye, G.Y.Li, B.-H.F.Juang,and K.Sivanesan,“Channel agnostic end-to-end learning based communication systems with conditional GAN,”in IEEE Globecom, pp.1–5,2018.
}
\cmt{downside of LocFree}On the other hand, the results in \cite{teganya2019location} show a promising gain in the power map accuracy especially in the scenarios with significant multi-path effects. 
Note, however, that the dimensionality of the feature space 
considered in \cite{teganya2019location} grows with the number of base stations in the covered area, \rev{and some locations may not receive signals from all base stations; this can lead to having data points with missing features, generating the need for dimensionality reduction and data completion techniques. } In opposition, LocB methods consider as their feature space the set of possible locations, which has constant dimensionality (2 or 3).



\subsection{Main Contributions}
  \begin{myitemize}

\myitem \cmt{precision of location information is valuable} 
The main idea of this work is to build a map estimation algorithm that can combine both the LocB and the LocF methods in a way that exploits the knowledge about location uncertainty. Notice that the more precise the location information of a node is, the more reliable the LocB map estimate is expected to be. \cmt{it would be great to have a support of this with a figure} An uncertainty measure regarding the location estimate acts, intuitively, as a measure of reliability of the LocB method (or expert) relative to the LocF method (or expert).
  
\myitem \cmt{Work hypothesis} It is demonstrated in this work that such an uncertainty estimate can be exploited to build more accurate CG maps. To this end, the proposed approach estimates the CG from:
a) the LocF features (e.g. center of mass (CoM) of the channel impulse response as in \cite{teganya2019location}), b) the estimated locations of the transmitter and receiver, and c) the location uncertainty information. 
\nextver{, while incorporating the fact that the uncertainty of the location estimates carries information about which of the two approaches should be trusted more. }
We postulate that the complementary benefits of the LocF and LocB estimators can be exploited efficiently by learning a LocB estimator and a LocF estimator, and combining their outputs with a gating function that incorporates the localization uncertainty. 
With higher uncertainty in the location estimation, the final estimate relies more on a LocF estimator; while otherwise, a LocB estimator has more weight on the final estimate. The weights of the LocB and LocF estimates are given by the gating function, which is optimized jointly with the experts or estimators.

The aforementioned way of estimating the CG is also attractive because of its simplicity, incorporating the divide-and-conquer philosophy of the mixture-of-experts (MoE) methods \cite{masoudnia2014mixture}.
This is suitable for our development because we seek an approach that enables estimating CGs at arbitrary locations where no sensors may be available, in addition to the sensor positions (although the accuracy of such estimates will be lower as compared to the accuracy obtained at the locations with sensor nodes). We expect that in real scenarios, some training points will have only one type of feature available. For the other training points with both types of features (LocB and LocF), the learning will be enhanced.

\rev{Moreover, this work is (to the best of our knowledge) the first to apply the LocF approach to the estimation of CG maps.}

\nextver{     \myitem \cmt{other ideas/showcase} It is interesting to demonstrate the power of spectrum cartography for instance by using real data, advanced generative models (such as REMCOM), advanced machine learning techniques/tricks, careful choice of kernel/architecture, diverse and challenging propagation scenarios(indoor). Also, recent work by Seung-Jun in multitask learning \cite{kim2019lifelong} has potential to be applied in spectrum cartography. }
    
    \end{myitemize}
The rest of the paper is structured as follows. In Sec.~\ref{sec:modeling}, our learning model is put forth in a MoE framework. In Sec.~\ref{sec:formulation}, the CG cartography problem is formulated. In Sec.~\ref{sec:optimization} a solution is derived, and the choice of hyper-parameters is later discussed (Sec.~\ref{sec:hpselection}). The results of numerical experiments are presented in Sec.~\ref{sec:experiments}, and conclusions are provided in Sec.~\ref{sec:conclusions}.

\section{Modeling}
    \label{sec:modeling}
Consider a transmitter located at $\bm x_{t} \in \mathcal{R}$, and a receiver located at $\bm x_{r}\in \mathcal{R}$, \rev{where $\mathcal{R}$ is the region of interest (typically a subset of $\mathbb{R}^2$ or $\mathbb{R}^3$)}. The CG between them is\nextver{modeled as a random variable} denoted by $C_{t,r}$ \rev{$\in \mathbb{R}$}. The main goal of CG cartography is to learn a function that can give a point estimate $\hat c_{t,r}$ of $C_{t,r}$, given the information gathered at each of the two terminals. \nextver{In this work, we assume that the information at each terminal consists of: features extracted from positioning pilot signals\acom{sent by base stations}, location estimates (obtained from the same pilot signals and the positions of the base stations), and the uncertainty regarding those location estimates. }
Specifically, let 
$\hat{\bm x}_{t}$ denote an estimate of the transmitter's location, and \rev{$e_t\in \mathbb{R}_+$} an uncertainty measure regarding $\hat{\bm x}_{t}$. Let $\bm \phi_t$\rev{$\in \mathbb{R}^M$} denote the vector containing the LocF features extracted from the pilot signals the same sensing node has received, \rev{where $M$ is the total number of features extracted} (more information about how $\bm \phi_t$ is obtained can be found in Sec. \ref{sec:locFreeFeatures}). 
A representation of the available information at the transmitter comes from stacking the aforementioned variables in the vector
$\bm \psi_t:=[ \bm \phi_t^\top, \; \hat {\bm x}_{t}^\top, \; e_t]^\top$ (analogously for the information at the receiver, $\bm \psi_r$).
\nextver{In a nutshell, the cartography approach we pursue in this paper consists in feeding the feature vectors of the $N$ recorded sensor locations to a machine learning model (such as a kernel machine or an NN) and using the produced model as a CG estimator.} 
The function we seek to learn is expressed in the form
$$
   \hat c_{t, r} =  f(\bm \psi_t, \bm \psi_r).
$$

\cmt{hybrid approach}    
With the twofold goal of estimating CG at arbitrary pairs of transmitter-receiver locations, and also where the sensing nodes do not have an accurate location estimate, our goal is to develop an approach such that for a query where $\bm \phi_t$ and $\bm \phi_r$ are available and $(\hat {\bm x}_t, \hat {\bm x}_r)$ have large uncertainty (or even are missing), one can leverage the LocF technique~\cite{teganya2019location}; and whenever $(\hat {\bm x}_t, \hat {\bm x}_r)$ are accurate and the features in $(\bm \phi_t, \bm \phi_r)$ are noisy, it should work similarly to LocB approaches such as \cite{kim2011cooperative,dall2011channel}. And between these two extreme situations, the idea discussed here is to combine both LocF and LocB estimates, exploiting the knowledge of uncertainty in the location information. 

\nextver{\cmt{desiderata: obtain uncertainty} As a secondary goal, the estimate can be enhanced by having a posterior distribution or, at least, a point estimate plus some uncertainty information such as an uncertainty region, or a covariance matrix. This constitutes uncertainty information, which is useful for a cognitive radio (CR) that intends to establish a spatial margin where the probability interference is low for primary users potentially located there.}

\nextver{If we want the function $f$ to make use of the LocB and the LocF estimates, as a first approach, one can consider a simple additive model:
    
$$f(\bm \psi_t, \bm \psi_r) = f_0 + f_l( \hat{\bm x }_t, \hat{\bm x }_r) + f_p(\bm \phi_t, \bm \phi_r)$$

where $f_0$ is an intercept term. However, this does not exploit any additional information about uncertainty in the localization estimation. 

If $f_p$ has zero mean, in a query point $\bm \psi_q:=[ \bm \phi_q^\top, \; \hat {\bm x}_{q}^\top]^\top$ where $\bm \phi_q$ is missing, the estimate of
$f(\bm \psi_q)$ is simply $f_0 + f_l(\hat{\bm x}_q)$; 
and similarly if $\hat{\bm x}_q$ is missing 
\cmt{Explain what is $f_0$}. 
More info regarding additive GPs: \cite{duvenaud2011additive, fox2012multiresolution}.
}

A popular approach to combine the estimates into $f$ is MoEs \cite{masoudnia2014mixture}. It is common practice to use a convex combination of the output of each of the experts, mainly because the combination coefficients can be interpreted as conditional probabilities of the events defining which of the experts has the best estimate for a given data point. This can be expressed without loss of generality by defining the gating function $\tilde g(\cdot)$:
\begin{equation}
\label{eq:generic_mixture}
    f(\bm \psi_t, \bm \psi_r) = \tilde g(\bm \psi_t, \bm \psi_r) f_l( \hat{\bm x }_t, \hat{\bm x }_r) + \big(1-\tilde g(\bm \psi_t, \bm \psi_r)\big) f_p(\bm \phi_t, \bm \phi_r).
\end{equation}
\nextver{The intercept $f_0$ is not present in this model as each of the two functions $f_l$ and $f_p$ can have its own intercept. }Such a gating mechanism is widespread in the ML literature not only because it is used in MoE, but also because of its presence in recurrent neural networks \nextver{(RNNs)}~\cite{chung2014empirical}. We postulate that incorporating the location uncertainty measure in the input of this gating function will result into an improved performance of the MoE-based CG map. This approach is justified because the LocB estimator is expected to perform better than the LocF one when the location estimate is sufficiently good. What "sufficiently good" means is something we expect the model will learn from the data. Moreover, the mixture allows each expert to focus its resources in learning its own part of the CG map in those areas where it is expected to perform better. The empirical results in Sec. \ref{sec:experiments} support this idea.


\subsection{Simple MoE Model: Gating as a Function of the Localization Uncertainty Measures}
\label{ss:gating}

The main idea in the model in \eqref{eq:generic_mixture} is to restrict each of the two experts in the mixture to have as input either location estimates, or LocF features. In order to keep the model complexity at the minimum, we propose to restrict the gating function to take as input only the uncertainty in estimating the locations $\hat{\bm x}_{t}, \hat{\bm x}_{r}$. This yields the simplest possible model that combines the aforementioned experts, and takes the location uncertainty into account. 

With $e_{\bm x,t}$ and $e_{\bm x,r}$  respectively denoting the uncertainty measures referring to estimating
$\hat{\bm x}_t$ and $\hat{\bm x}_r$, we will design a gating function 
\begin{align}
g: \mathbb{R}_+^2 &\rightarrow [0,\,  1] 
\end{align}
that takes the localization error vector
$\bm e:=[e_{\bm x,t},\; e_{\bm x,r}]^\top$ as an input for any transmitter-receiver pair $(t,r)$. \acom{R1: The gating function takes as input only the uncertainty in estimating the locations. The uncertainty of the LocF feature is not considered} \rev{A more sophisticated model could also incorporate the uncertainty associated with $(\bm \phi_t, \bm \phi_r)$, but it is not clear whether the gain in performance would be significant.} The MoE model can be written as:
\begin{equation} 
\label{eq:model}
f(\bm \psi_t, \bm \psi_r) =      
     g( \bm e )  f_l( \hat{\bm x}_t, \hat{\bm x}_r)
     + \left(1-g( \bm e )\right) f_p( \bm    \phi_t,    \bm \phi_r) 
\end{equation}
For this model, it is clear that $g(\bm e )$ should give less emphasis on $f_l(  \hat{\bm x}_t,\hat{\bm x}_r)$ when either $ e_{\bm x,t} $ or $ e_{\bm x,r} $ is large. 

Successful learning of the hybrid (MoE) model $f(\bm \psi)$ entails some advantages: \begin{myitemize}
    \myitem The information carried by the location uncertainty measure allows to give as much weight as is needed to the LocB and LocF estimates.
    \myitem 
    Whenever the location estimates are deemed reliable, the MoE gives more weight to the location-based estimate, which mitigates the relative difficulty of the location-free estimation to generalize due to the higher dimensionality of the positioning features \acom{have we already mentioned the issue of dimensionality with LocF?}. To see this, consider the case where two different queries are performed for the same Tx-Rx pair, but the positioning (pilot) signals are received from different location sources. While a pure localization-free approach might fail to generalize, the CG can still be estimated successfully if the localization algorithm identifies the location correctly. 
    \myitem 
    One can even evaluate the CG map for an arbitrary pair of locations when there is no sensing node at either one or both locations. In such a case, the estimate is simply given by $f_l(\bm x_t, \bm x_r)$. \acom{maybe necesssary to adjust the weight to e.g. $g(0)$}  
\end{myitemize} 
\cmt{choice of $g(\cdot)$} \nextver{Let $\mathcal{G}$ denote the set of feasible instances of $g(\cdot)$. } 

The gating function should be component-wise non-increasing, i.e., \acom{, $$ 
        e_{\bm x,t} \geq e^\prime_{\bm x,t} \, \wedge \, 
        e_{\bm x,r} \geq e^\prime_{\bm x,r}
    \longrightarrow 
        g( [e_{\bm x,t},\; e_{\bm x,r}]^\top  ) \leq 
        g( [e^\prime_{\bm x,r},\; e^\prime_{\bm x,t}]^\top),
    $$ 
    which can be rewritten compactly as}
    $$
        g(\bm e) \leq g(\bm e^\prime)\; \forall \; (\bm e, \bm e^\prime) \text{ such that } \bm e \succeq \bm e^\prime
    $$
    where \notation{the notation $\bm a \succeq \bm b$ denotes for $\bm a, \bm b \in \mathbb{R}^N$ that $[\bm a]_i \geq [\bm b]_i \; \forall i \in [1, N]$}. 
    \begin{myitemize}%
    \myitem Under the assumption of symmetric channels, $g( \bm e )$ should also be symmetric, i.e. 
    $$
        g( [e_{\bm x,t},\; e_{\bm x,r}]^\top  ) = 
        g( [e_{\bm x,r},\; e_{\bm x,t}]^\top); \; 
        \forall \; \bm e \in \mathbb{R}_+^2. 
    $$
    \acom{already said before (3): \myitem and it must lead to a convex combination of $f_l$ and $f_p$, i.e., $g(\bm e) \in [0, 1], \;\; \forall \bm e \, \succeq 0$.}
    \myitem%
    For large $\|\bm e\|$, meaning an unknown location, the weight given by the gating function to the LocB estimator should vanish: $\lim_{\|\bm e\|\to\infty} g(\bm e) =0$. However, it is not necessary to force $ g(\bm 0) = 1$, as the LocB map is imperfect and a certain contribution from the LocF one may give better performance, even if some locations are deemed perfectly known. \nextver{Forcing $ g(\bm 0) = 1$ may have undesirable effects if none of the training samples has a perfect location estimate associated with it, and may make the function $f_l(\cdot)$ too large when used for a pair of locations which are known perfectly.}
\end{myitemize}  
\nextver{For simplicity, one could also assume an additive form: $ g( \bm e ) := \big(\tilde \sigma(e_{\bm x,r}) + \tilde \sigma(e_{\bm x,t})\big)/2$. \acom{Luismi: I expect the additive form to have worse performance. I implemented it in Matlab too, because it was easy. We can compare two schemes, each based on the two kinds of gating function}}


\subsection{Positioning Signal-Based Features}
\label{sec:locFreeFeatures}
\cmt{Center of mass} The LocF features extracted from the pilot signals that will be used in this work are the CoMs described in \cite{teganya2019location}. Let $CoM_{m, n}$ denote the center of mass of the cross-correlation between the pilot received at the $n$-th sensing node from the $m$-th positioning signal source, and the pilot received from the reference source (which is arbitrary and the same for all sensing nodes). The feature vector is then $\bm \phi_n:= [CoM_{1, n},\ldots,CoM_{M,n}]$. 
The reason of the choice of this kind of features is that they evolve smoothly over space and are robust to the pilot distortions caused by multipath. Therefore, the function $f_p$ can be easily learnt with such features. 
\nextver{Possible extension: Adding uncertainty measures for positioning pilot signal-based features}

When the region to map is large, it is likely that some of the base stations that send pilot signals are so far away from a given location that some features become missing, either in some of the training points, or in query points. While a few missing TDoAs is not a big issue for a localization algorithm as long as enough sources are visible, the way the LocF map is designed requires all entries in the query feature vector to carry values. The technique for imputing such missing features in \cite{teganya2019location} can be seamlessly applied in the application in this work. In a nuthshell, this technique is based in the assumption that the LocF features lie in a low-dimensionality subspace. Such a subspace is learnt from the training data using a low-rank matrix completion technique. \nextver{To impute missing values in the training set, matrix completion schemes are used that exploit prior information on the subspace where the features lie. To impute missing features in query/test locations, one has to find the closest point in the subspace where the training points were deemed to lie close to.} \nextver{Advanced kernel formulations such as hierarchical kernel learning are robust to missing features and will be considered in future work.}

\subsection{Location Estimates and Uncertainty Measures}

Location estimates can be obtained in practice by extracting TDoA measurements from the pilot signals \cite{el2013low}, and solving the localization problem e.g. along the lines of \cite{wang2016robust}. It is assumed that the location estimator also gives a scalar measure of the uncertainty of the location estimate. This measure can be, e.g. the spectral radius of a covariance matrix, or the diameter of an uncertainty region for a given level of confidence. The procedure for obtaining such an uncertainty measure is left out of the scope of the present paper. Our experiments will rely on synthetically-generated uncertainty measures.

\acom{For a first proof of concept, we will use synthetic location estimates, produced by adding Gaussian noise to the true locations.}

\nextver{In a further, more realistic step, we plan to use the Robust localization under NLOS measurements proposed by Wang et al \cite{wang2016robust}.
    
\subsection{Uncertainty measures for localization}
A point estimate of a sensing node's location is given by the minimizer of a cost function. However, the shape of the localization problem's objective function around its minimizer could carry information about the uncertainty in the estimate. 
 
 \begin{myitemize}
     \myitem\cmt{Hessian} In simple gaussian estimation problems, the Hessian informs about the variance of the estimate. Intuitively, the "sharper" the cost function, the more accurate the localization will be. A measure of the Hessian evaluated at the optimizer, such as the spectral radius $\rho(\bm H)$ is expected to be proportional to the uncertainty.
     \myitem\cmt{Approximate Hessian} In more involved problems, the Hessian may be variant or even not exist at the optimum (nonsmooth objective function). In such a case, one can try with a sample-based estimate of the Hessian. This is cheap to obtain, and comes at almost zero cost if the localization was solved via grid search.
     \myitem\cmt{Geometry} If we can obtain an uncertainty region from a localization algorithm, its diameter is also informative of the uncertainty in the localization.
     \myitem\cmt{Level set} A level set of the objective function is a simple approximation of an uncertainty region. An open question in this case is how to establish the level, e.g. minimum value plus a constant.
\end{myitemize}
}

\section{Problem Formulation}
\label{sec:formulation}
\newcommand{\Np}{{N_p}}

With $\Np$ denoting the number of training transmitter-receiver pairs, let $t(n)$ and $r(n)$ respectively denote the indices of the transmitter and the receiver of the $n$-th pair; and let $\tilde c_n$ denote the measured CG between them (i.e., a noisy observation of $C_{t(n), r(n)}$). Adopting a regularized least-squares criterion, the CG map training can be expressed as:
\begin{equation}
    \underset{
        {f}\in \mathcal{F}
    }
    {\text{minimize}} \quad 
    \frac{1}{\Np}\sum_{n=1}^\Np  \left(
        \tilde{c}_n
        - f(\bm \psi_{t(n)}, \bm \psi_{r(n)})
        \right)^2 
    + \lambda \tilde \Omega(f)
\end{equation}
One valid approach is to define a neural network (NN) architecture, letting $\mathcal{F}$ denote the set of all functions that NN can express, and defining $\tilde \Omega$ as a regularizer that depends on the NN weights. However, the number of training samples for such an NN to achieve good generalization may be far beyond the number of samples available in a practical case.

We aim at learning the function $f$ in a structured way, by using the MoE described in Sec. \ref{sec:modeling}. We expect the number of samples needed for good generalization to be much smaller than that with a generic model.

The joint optimization of the experts and the gating function is written as the regularized functional estimation problem \eqref{eq:explicit_problem} at the bottom of the page,
\begin{figure*}[b]    
    \begin{align}
    \label{eq:explicit_problem}
    \underset{
        {f_p}\in \mathcal{F}_p, 
        {f_l}\in \mathcal{F}_l, 
        g \in \mathcal{G}
    } 
    {\text{minimize}} \quad &
    \textstyle
    \frac{1}{\Np}\sum_{n=1}^\Np  \left(
        \tilde{c}_n
        - g( \bm e_n )     f_l( \hat{\bm x  }_{t(n)},  \hat{\bm x  }_{r(n)})
        -(1- g( \bm e_n )) f_p(\bm \phi_{t(n)}, \bm \phi_{r(n)}) 
    \right)^2 \nonumber
    \\
    & 
    \textstyle+ \left(\frac{1}{\Np}\sum_{n=1}^\Np g( \bm e_n )\right)\lambda_l \Omega( f_l) 
    \textstyle+ \left(\frac{1}{\Np}\sum_{n=1}^\Np 1 - g( \bm e_n )\right)\lambda_p \Omega( f_p), 
\end{align} 
\end{figure*}
where 
$\mathcal{G}$ denotes the set of instances of $g(\cdot)$ that have the properties discussed at the end of Sec. \ref{ss:gating}, and
$\mathcal{F}_p$ and $\mathcal{F}_l$ are model-specific function spaces such as a reproducing kernel Hilbert space (RKHS) for a given kernel, or a set of functions implemented by an NN. The terms in parentheses multiplying $\lambda_p$ and $\lambda_l$ are intended for balancing the contribution of the regularization terms for any value of $g(\bm e_n)$. If these terms were absent, many algorithmic attempts to solve this problem would very likely fall into one of the two trivial solutions, namely: $g(\bm e) = 0 \; \forall \bm e$ or $g(\bm e) = 1 \; \forall \bm e$, which respectively imply $f(\bm \psi_n) = f_p(\bm\phi_{t(n)}, \bm\phi_{r(n)})$, or $f(\bm \psi_n) = f_l(\hat{\bm x}_{t(n)}, \hat{\bm x}_{r(n)})$. \nextver{More information will be given after the proposed algorithm is presented.}
The problem of estimating the coefficients of a set of kernel machines whose outputs are combined using a given gating function is presented and discussed in \cite{santarcangelo2015kernel}. Differently, the joint optimization of the experts and the gating function is done in a novel way here, exploiting the problem structure to yield a low-complexity algorithm. \acom{(If we do it successfully, we can claim that this is one of the main ML-related contributions of our paper.)}%
\acom{This problem can be rewritten in vector form as:
\begin{equation} \label{eq:generic_fe_vector} 
    \begin{split} 
        \underset{
            {f_p}\in \mathcal{F}_p, 
            {f_l}\in \mathcal{F}_l, 
            g \in \mathcal{G}
        } 
    {\text{minimize}} \quad &
    \frac{1}{\Np}\left \Vert \tilde{\bm{c}}- 
        \begin{bmatrix}
            1-g(\bm e_1) & & \\ 
            & \ddots & \\ 
            & & 1-g(\bm e_\Np)
        \end{bmatrix} 
        \bm{f}_p - 
        \begin{bmatrix}
            g(\bm e_1) & & \\ 
            & \ddots & \\
            & &g(\bm e_\Np)
        \end{bmatrix} 
        \bm{f}_l\right \Vert ^2 
        + \lambda_p \Omega( f_p)+  \lambda_l \Omega( f_l)
    \end{split} 
\end{equation}
or, more compactly,
\begin{equation}  \label{eq:generic_fe_vector} 
    \begin{split} 
        \underset{f_p, f_l, g} {\text{minimize}}  \quad 
        & \frac{1}{\Np}\left \Vert \tilde{\bm{c}}- 
        (\bm I - \bm D)
        \bm{f}_p - 
        \bm D
        \bm{f}_l\right \Vert ^2 
        + \lambda_p \Omega( f_p)+  \lambda_l \Omega( f_l),
    \end{split} 
\end{equation}
where }Upon scaling up the objective by the constant $N_p$, and defining
\begin{subequations}
\begin{align}   
    \bm{f}_l := &[ 
    f_l(\hat{\bm x }_{r(1)}, \hat{\bm x }_{r(1)}), 
    \dots, 
    f_l( \hat{\bm x }_{t(N_p)}, \hat{\bm x }_{r(N_p)})]^\top, 
    \\
    \bm{f}_p := &[ 
    f_p(\bm \phi_{t(1)}, \bm \phi_{r(1)}), 
    \dots, 
    f_p(\bm \phi_{t(N_p)}, \bm \phi_{r(N_p)})]^\top, 
    \\
    \bm g := & 
    \left[ 
        g(\bm e_1), g(\bm e_2), \dots g(\bm e_\Np) 
    \right]^\top,
    \label{eq:v_g}
\end{align}
\end{subequations}
problem \eqref{eq:explicit_problem} can be rewritten equivalently as
\newcommand{\ewm}{\odot}
\begin{align}  \label{eq:generic_fe_vector} 
    \begin{split} \hspace{-2mm}
        \underset{
        {f_l}\in \mathcal{F}_l, 
        {f_p}\in \mathcal{F}_p, 
        g \in \mathcal{G}
    } 
    {\text{minimize}} \quad &
    \left \Vert 
        \tilde{\bm{c}}
        -  \bm g \ewm        \bm{f}_l
        - (\bm 1 - \bm g) \ewm \bm{f}_p      
        \right \Vert ^2 \\
        + & \lambda_l \bm 1^\top \bm g \, \Omega( f_l)
        + \lambda_p \bm 1^\top (\bm 1 - \bm g)\Omega( f_p),
    \end{split} 
\end{align}
where \notation{$\ewm$ denotes element-wise vector (Hadamard) product}.

\section{Optimization}
\label{sec:optimization}

At this point, the functions $f_p, f_l, g$ can be learnt using several different approaches. \begin{myitemize}%
\myitem\cmt{back-prop}%
If such functions are expressed parametrically, one can compute (automatically via back propagation) the gradient of the cost function in \eqref{eq:generic_fe_vector} \acom{with respect of the parameters}, and run a gradient-based minimization algorithm to seek a local minimum (as is the common practice for NNs).

\myitem\cmt{block-coordinate} An alternative approach \nextver{that does not require to define the functions parametrically} is to solve the problem in \eqref{eq:generic_fe_vector} using block-coordinate minimization (BCM):

\acom{old version with D matrix \begin{subequations} \label{eq:bcd}
\begin{align}
f_l^{(k+1)}    := &\arg \min_{f_l \in \mathcal{F}_l}  
\left \Vert \tilde{\bm{c}}- 
        (\bm I - \bm D^{(k)}) \bm{f}^{(k+1)}_p - \bm D^{(k)} \bm{f}_l\right \Vert ^2
        + \lambda_l 1^\top \bm g \, \Omega( f_l) \label{eq:bcd_l}
\\
f_p^{(k+1)}    := &\arg \min_{f_p \in \mathcal{F}_p}  
\left \Vert \tilde{\bm{c}}- 
        (\bm I - \bm D^{(k)}) \bm{f}_p - \bm D^{(k)} \bm{f}_l^{(k)}\right \Vert ^2 
        + \lambda_p \bm  1^\top (\bm 1 -\bm g) \Omega( f_p) \label{eq:bcd_p}
\\
g^{(k+1)} := &\arg \min_{g \in \mathcal{G}}  \left \Vert \tilde{\bm{c}}- 
        (\bm I - \bm D) \bm{f}^{(k+1)}_p - \bm D \bm{f}^{(k+1)}_l\right \Vert ^2
        \label{eq:bcd_g}
\end{align}
\end{subequations}}

\begin{subequations} \label{eq:bcd}
\begin{align}
f_l^{(k+1)}    := \arg &\min_{f_l \in \mathcal{F}_l} 
\left \Vert \tilde{\bm{c}}- 
        (\bm 1 - \bm g^{(k)}) \ewm \bm{f}^{(k)}_p - \bm g^{(k)} \ewm \bm{f}_l\right \Vert ^2
        \nonumber \\
        & + \lambda_l 1^\top \bm g \, \Omega( f_l) \label{eq:bcd_l}
\\
f_p^{(k+1)}    := \arg &\min_{f_p \in \mathcal{F}_p}  
\left \Vert \tilde{\bm{c}}- 
        (\bm 1 - \bm g^{(k)}) \ewm \bm{f}_p - \bm g^{(k)} \ewm \bm{f}_l^{(k+1)}\right \Vert ^2 
        \nonumber \\
        & + \lambda_p \bm  1^\top (\bm 1 -\bm g) \Omega( f_p) \label{eq:bcd_p}
\\
g^{(k+1)} := \arg &\min_{g \in \mathcal{G}} \left \Vert\tilde{\bm{c}}- \bm{f}^{(k+1)}_p
          - (\bm{f}^{(k+1)}_p-\bm{f}^{(k+1)}_l )
          \ewm \bm g \right \Vert  \nonumber
          \\
          & + \left(
             \lambda_l \Omega(\bm f_l^{(k+1)})
            -\lambda_p \Omega(\bm f_p^{(k+1)})
            \right)\bm 1^\top \bm g
        \label{eq:bcd_g}
\end{align}
\end{subequations}
BCM converges monotonically to a local minimum of \eqref{eq:generic_fe_vector}.
\nextver{\myitem\cmt{bcd} If we implement the solution of the above problems via a gradient algorithm, we can also try block-coordinate descent (BCD) alternating gradient steps for each of the three problems one at a time.}
\end{myitemize}

Whenever $\mathcal{F}_l$ is an RKHS (denoted by $\mathcal{H}_l$) with associated kernel $\kappa_l(\cdot, \cdot)$, and the regularizer $\Omega$ is the associated RKHS norm $\| \cdot \|_{\mathcal{H}_l}^2$, the subproblem \eqref{eq:bcd_l} is a standard kernel ridge regression (KRR) problem; the same applies to \eqref{eq:bcd_p}.
According to the Representer Theorem\cite{scholkopf2001representer}, there exist minimizers for \eqref{eq:bcd_l} and \eqref{eq:bcd_p} with the following forms, respectively:
\begin{subequations}
    \begin{align}
        f_l(\hat{\bm x  }_t, \hat{\bm x  }_r)=
        & \sum_{n=1}^{N_p} \alpha_{l,n}\kappa_l([ \hat{\bm x  }_t^\top, \; \hat{\bm x  }_r^\top]^\top, [\hat{\bm x }_{t(n)}^\top, \; \hat{\bm x  }_{r(n)}^\top]^\top)
        \\
        f_p(\bm \phi_t, \bm \phi_r)= 
        & \sum_{n=1}^{N_p} \alpha_{p,n}\kappa_p([ \bm \phi_t^\top, \; \bm \phi_r^\top]^\top, [\bm \phi_{t(n)}^\top, \; \bm \phi_{r(n)}^\top]^\top).  
    \end{align}
\end{subequations}
and if we define the kernel matrix $\bm K_l$ such that $[\bm K_l]_{ij} = \kappa_l([ \hat{\bm x  }_{t(i)}^\top, \; \hat{\bm x  }_{r(i)}^\top]^\top, [\hat{\bm x }_{t(j)}^\top, \; \hat{\bm x  }_{r(j)}^\top]^\top)$, 
define $\bm K_p$ analogously, and $\bm D^{(k)} \triangleq \mathrm{Diag}(\bm g^{(k)})$, then solving (\ref{eq:bcd_l}-\ref{eq:bcd_p}) boils down to:
\acom{ previous ver
\begin{subequations} \label{eq:kernel_separatexxx}
\begin{align}
\bm\alpha_p^{(k+1)}    := \arg \min_{\bm \alpha_p} & 
    \frac{1}{\Np}\left \Vert \tilde{\bm{c}}- 
        (\bm I - \bm D^{(k)}) 
        \bm{K}_p\bm \alpha_p - 
        \bm D^{(k)} 
        \bm{K}_l\bm \alpha_l^{(k)}\right \Vert ^2 + 
        \lambda_p \bm{\alpha}_p^{\top}\bm{K}_p\bm{\alpha}_p \label{eq:bcd_krr_1}
\\
\bm\alpha_l^{(k+1)}    := \arg \min_{\bm \alpha_l} & 
    \frac{1}{\Np}\left \Vert \tilde{\bm{c}}- 
        (\bm I_N - \bm D^{(k)}) 
        \bm{K}_p \bm \alpha_p^{(k+1)} - 
        \bm D^{(k)} 
        \bm{K}_l\bm \alpha_l \right \Vert ^2  +
        \lambda_l \bm{\alpha}_l^{\top}\bm{K}_l\bm{\alpha}_l \label{eq:bcd_krr_2}
\end{align}
\end{subequations}
}
\begin{subequations} \label{eq:kernel_separate}
\begin{align}
\bm\alpha_l^{(k+1)}    := \arg \min_{\bm \alpha_l} & 
    \left \Vert \tilde{\bm{c}}- 
        (\bm I - \bm D^{(k)}) 
        \bm{K}_p \bm \alpha_p^{(k)} - 
        \bm D^{(k)} 
        \bm{K}_l\bm \alpha_l \right \Vert ^2  \nonumber \\
        & +
        \lambda_l \bm 1^\top \bm g \,
        \bm{\alpha}_l^{\top}\bm{K}_l\bm{\alpha}_l \label{eq:bcd_krr_l}
        \\
\bm\alpha_p^{(k+1)}    := \arg \min_{\bm \alpha_p} & 
    \left \Vert \tilde{\bm{c}}- 
        (\bm I - \bm D^{(k)}) 
        \bm{K}_p\bm \alpha_p - 
        \bm D^{(k)} 
        \bm{K}_l\bm \alpha_l^{(k+1)}\right \Vert ^2 \nonumber \\ 
        & + 
        \lambda_p \bm 1^\top (\bm 1 - \bm g) \, \bm{\alpha}_p^{\top}\bm{K}_p\bm{\alpha}_p \label{eq:bcd_krr_p}
\end{align}
\end{subequations}
and it holds that $\bm f_p = \bm{K}_p\bm{\alpha}_p$, and $\bm f_l = \bm{K}_l\bm{\alpha}_l$.
\nextver{The equations above can be solved in closed form as:
\begin{subequations}
\begin{align}
\bm\alpha_p^{(k+1)}  & := \left(
        (\bm I_\Np - \bm D^{(k)})\bm{K}_p
        +\lambda_p \bm 1^\top (\bm 1 - \bm g) \bm I_\Np
    \right)^{-1} 
    \left( 
        \tilde{\bm{c}}-  
        \bm D^{(k)} \bm{K}_l
        \bm \alpha_l^{(k)}
    \right)
\\
\bm\alpha_l^{(k+1)}  & := \left(
        \bm D^{(k)} \bm{K}_l\bm 
        +\lambda_l \bm 1^\top \bm g \bm I_\Np
    \right)^{-1} 
    \left( 
        \tilde{\bm{c}}-  
        (\bm I_\Np - \bm D^{(k)})\bm{K}_p 
        \bm \alpha_p^{(k+1)}
    \right)
\end{align}
\end{subequations}
}
\nextver{For the particular case when both the LocB and the LocF estimators can be optimized jointly, the iterations (\ref{eq:bcd_l}-\ref{eq:bcd_p}) can be replaced with \eqref{eq:bcd_twoStep_1}, shown at the bottom of the page.
\begin{figure*}[b]
\begin{align} 
\{f_p^{(k+1)}, f_l^{(k+1)}\}    := \arg \min_{f_p \in \mathcal{F}_p, f_l \in \mathcal{F}_l}  \quad &
    \left \Vert 
        \tilde{\bm{c}}
        -  \bm g^{(k)} \ewm        \bm{f}_l
        - (\bm 1 - \bm g^{(k)}) \ewm \bm{f}_p      
        \right \Vert ^2 \nonumber
        \\
        &+ \lambda_l \bm 1^\top \bm g^{(k)} \, \Omega( f_l)
        + \lambda_p \bm 1^\top (\bm 1 - \bm g^{(k)})\Omega( f_p). 
        \label{eq:bcd_twoStep_1}
\end{align}
\end{figure*}
}

When both $\mathcal{F}_l$ \emph{and} $\mathcal{F}_p$ are RKHSs, \eqref{eq:kernel_separate} can be substituted with a joint optimization whose closed form is \eqref{eq:optweights} (shown at the top of next page). It turns out that a related model is proposed and discussed in \cite{santarcangelo2015kernel}, but the gating function there is a generic (softmax) function, which would make the optimization in \eqref{eq:bcd_g} nonconvex. An alternative approach is proposed in this paper, based on exploiting the structure of the problem at hand to design a low-complexity solver.
\newcommand{\Id}{\bm I_\Np}
\begin{figure*}[t]
\begin{equation}\label{eq:optweights}
    \left[ 
        \begin{matrix} 
            \bm\alpha_p^{(k+1)} \\
            \bm\alpha_l^{(k+1)}
        \end{matrix} 
    \right]
    =
    \left[ 
        \begin{matrix} 
            (\Id - \bm D^{(k)})^2\bm{K}_p+\lambda_p \bm 1^\top (\bm 1 - \bm g) \Id
            & (\Id- \bm D^{(k)}) \bm D^{(k)} \bm{K}_l
            \\ 
            \bm D^{(k)} (\Id- \bm D^{(k)}) \bm{K}_p
            & (\bm D^{(k)})^2 \bm{K}_l+\lambda_l \bm 1^\top \bm g \Id
        \end{matrix}
    \right]^{-1} 
    \left[
        \begin{matrix} 
            (\Id- \bm D^{(k)})\tilde{\bm{c}}  \\
            \bm D^{(k)} \tilde{\bm{c}} 
        \end{matrix} 
    \right]
\end{equation}
\end{figure*}




\nextver{
\cmt{solving with respect to $\bm D$} 
Regarding the optimization of $g(\cdot)$, two approaches can be taken:
\begin{myitemize}
    \myitem\cmt{Parametric function} 
    One option is to let $g(\cdot)$ be a parametric function, define $\mathcal{G}:=\{g(\cdot; \theta) \vert \theta \in \Theta\}$, and optimize over $\theta$. 
    \acom{ a parametric form for $g(\cdot)$, i.e.,  $g(\bm e; \theta)$ and formulate the parameter estimation problem as 
    \begin{equation}
        \underset{\bm \theta} {\text{minimize}} \quad \left \Vert  \tilde{\bm{c}}- \bm{f}_p- \begin{bmatrix}
            g(\bm e_1; \theta) & & \\ 
            & \ddots & \\
            & &g(\bm e_\Np; \theta)
        \end{bmatrix} 
        (\bm{f}_l - \bm{f}_p) \right \Vert^2
    \end{equation}}
Note that for a model as simple as a softmax, the optimization becomes non-convex. For more complex architectures such as NNs, the model selection for $g(\cdot)$ increases significantly the hyperparameter search space.
\end{myitemize}
    }
 \cmt{Nonparametric monotonic function} 
Recall that in the RKHS case, (\ref{eq:bcd_l}-\ref{eq:bcd_p}) become convex problems. If the optimization over $g(\cdot)$ is also formulated as a convex problem, one can expect much more efficient learning. In fact, one can directly incorporate the properties of $g(\cdot)$ described in Sec. \ref{ss:gating} 
in the definition of $\mathcal{G}$, so that \eqref{eq:bcd_g} becomes:
    \begin{subequations}\label{eq:solve_for_g}
    \begin{align}      
       \underset{\bm g \in \mathbb{R}^\Np} {\text{minimize}} \quad &
       \left \Vert  \tilde{\bm{c}}- \bm{f}_p^{(k+1)} - \mathrm{Diag}(\bm{f}_l^{(k+1)} - \bm{f}_p^{(k+1)}) \bm g \right \Vert^2
       \nonumber \\
        & + \left(\lambda_l \Omega(\bm f_l^{(k+1)})
            -\lambda_p \Omega(\bm f_p^{(k+1)})
            \right)\bm 1^\top \bm g
       \\
       \text{s. to:} \quad &
            \bm 0 \preceq \bm g \preceq \bm 1
            \\
            & [g]_{i} \leq [g]_{j} \; \forall \; (i, j) \;\;
            \mathrm{s.t.}\; \bm e_i \succeq \bm e_j\; 
       \label{eq:many_constraints}
    \end{align}\end{subequations} 
    which is a standard convex quadratic problem with affine constraints. Regarding symmetry, it can be enforced easily (not only on the gating function but also on $f_p$ and $f_l$) by augmenting the training set, i.e., for each sample $(\tilde c_n, \bm \psi_{t,n}, \bm \psi_{r,n})$, adding its counterpart $(\tilde c_n, \bm \psi_{r,n}, \bm \psi_{t,n})$ to the training set. Once $\bm g$ is found, any gating function $g(\cdot)$ in agreement with \eqref{eq:v_g}
    will be optimal for the training set for fixed $\bm f_p, \bm f_l$.
    \acom{There is no need to solve for the function $g$ during training, as the only values that are needed are those in the vector $\bm g$.}
    Once the overall procedure has converged, an instance of $g(\cdot)$ can be recovered easily by interpolating the values in $\bm g$ with a suitable interpolation technique (e.g., $K$ nearest neighbors (KNN)).
   
\textbf{Remark.} The collection of constraints in \eqref{eq:many_constraints} is written with as many constraints as partial order relations in the set $\{\bm e_n\}_{n=1}^\Np$, for clarity. The number of constraints grows superlinarly with $N_p$. To avoid excess of complexity, the number of constraints can be reduced by building a directed acyclic graph (DAG) with the latter order relations, and computing its transitive reduction. This results into a DAG encoding the minimal set of constraints (that implies all others by the transitive property), yielding an equivalent problem with much fewer constraints.
   
   \nextver{\cmt{Challenges} The performance of the hybrid estimator depends on the quality of the interpolation implemented after determining the vector $\bm g$. This model enjoys almost absolute freedom on the choice of the function $g(\cdot)$, but can yield highly non-smooth functions when the number of training samples is small. This could lead to overfitting, and suggests the possibility of controlling for the smoothness of $g(\cdot)$ with a regularizer that penalizes its roughness. This will be studied in the journal version.
   }
   \nextver{\cmt{dropout}An alternative to defining a regularizing term for the g function is to effect regularization via dropout. This would be implemented by ignoring a random subset of the summands in \eqref{eq:rkhs_functions} in each iteration. We will possibly need to switch from BCM to projected gradient descent to optimize the g function with dropout.}

 
\subsection{Hyperparameter Selection}
\label{sec:hpselection}

If a Gaussian/RBF kernel is used, the kernel functions $(\kappa_l, \kappa_p)$ have width parameters $\sigma_l$ and  $\sigma_p$. The proposed estimator has then the following hyperparameters: $\lambda_l, \lambda_p, \sigma_l, \sigma_p$. It may be challenging to adjust all these hyperparameters by grid search and cross-validation (CV), for two reasons: a) the dimensionality of the search space is 4, as opposed to the search space for LocB or LocF which is 2; and b) the computation required to train the MoE is much higher than that for each of the experts separately, because of the iterative loop and the relatively slow convergence of BCM. A simplified procedure is proposed, based on selecting the hyperparameters which are CV-optimal for the LocB and LocF estimators separately, and then reusing the same hyperparameters for the MoE. The procedure is tabulated as Alg. \ref{alg:hyper-and-train}.

\rev{The computational cost of Alg.  \ref{alg:hyper-and-train} depends on: a) the number of elements in the grids where $\lambda_p$ and $\lambda_l$ are searched; b) the number of training samples $N_p$; and c) the number of iterations required for the for loop in Step 5 to converge. The dominating step with a practical configuration is Step 6, whose complexity is $\mathcal{O}(N_p^3)$ due to the matrix inversion in \eqref{eq:optweights}.}

\begin{algorithm}
    \caption{
        Hyper-parameter selection and MoE training
        }
    \label{alg:hyper-and-train}
    \textbf{Input:} 
        Training data $\{\bm \psi_{t,n}, \bm \psi_{r,n}, \tilde c_n\}_{n=1}^{N}$ \\
    \textbf{Output:} 
        CG estimating function $f(\bm \psi_t, \bm \psi_r)$
    \begin{algorithmic}[1] 
        \State{Select hyperparameters $(\lambda_p, \sigma_p)$ for the LocF CG estimator via CV and grid search}
        \State{Select hyperparameters $(\lambda_l, \sigma_l)$for the LocB CG estimator via CV and grid search}
        \State{Initialize $g(\bm e) = 1/2 \, \forall \bm e$ by defining $\bm g = \bm 1 / 2$}
        \State{Set $(\lambda_l, \lambda_p, \sigma_l, \sigma_p)$ as hyperparameters for the MoE}
        \For {$k=1, 2, \ldots $} (until convergence) 
            \State{Joint KRR coefficients optimization via \eqref{eq:optweights}}
            \State{Optimize gating function via \eqref{eq:solve_for_g}}
        \EndFor
        \State{\textbf{Return} $f(\bm \psi_t, \bm \psi_r)$ via \eqref{eq:model} }
    \end{algorithmic}
\end{algorithm}%
     
\nextver{

\section{Some comments towards analysis}   
Clearly, the function $f_p$ has a high-dimensional domain (as many dimensions as base stations/pairs), whereas $f_l$ has a low- (only 2- or 3-) dimensional domain. Moreover, $f_p$ is understood as a high-resolution function (probably hard to track for all variables), whereas $f_l$ is a low-resolution function.

The case of $f_l$ and the tracking of its time dynamics is covered by~\cite{kim2011cooperative,dall2011channel}.

\textbf{Question.} Can we say something about the identifiability of these 2 functions that are complementary to each other?

\cmt{Convergence} Convergence to the global optimum? Convergence to stationary point? \cmt{Balta \ra theorems}
}
\section{Experiments}
\label{sec:experiments}
A wireless propagation environment is simulated using an adapted version of the ray-tracing software in~\cite{hosseinzadeh2017enhanced}. The original code considers a set of walls and several sources to generate a power map accounting for direct, first, and second order reflected paths. \nextver{over a structure located in the area of interest. This structure comprises parallel vertical planes that model the walls of a building and where some of the sources are positioned inside, others outside the building.} The original source code has been modified to generate CGs between any two points in the area where the set of walls lie.

A set of positioning sources (e.g., base stations) are also simulated, and their pilot signals are transmitted through the aforementioned environment, so the received pilots are affected by the same multipath and attenuation that creates the CGs. The features to be used by the LocF estimators are obtained as the CoM of the cross-correlation between each pair of localization source pilots [cf. Sec. \ref{sec:locFreeFeatures}].
The location estimates $\hat {\bm x}$ to be used by the LocB estimator are generated synthetically  
by adding random noise $\sim \mathcal{N}(\bm 0, \sigma_x^2 \bm I)$ to the true locations of the simulated nodes. The location uncertainties $e_{x}$ are also synthetically generated by adding random noise $\sim \mathcal{N}(0, \sigma_e^2)$ to the Euclidian distance between the true location and its estimate. For these experiments, $\sigma_x:=7$ m, and $\sigma_e:=0.3 $ m (so that the uncertainty is significant and its measure is consistent with the deviation of the estimate from the true location). 
 \nextver{The positioning sources are labeled arbitrarily, and reference source for TDoA measurements is located at $(x, y) = (20, 10)$. For each training sample, a receiving user equipment (UE) terminal is simulated at its corresponding location, and the pilot signals are received from the base stations. Each TDoA is estimated by finding the peak value of the cross-correlation between the pilot signal from each source and that from the reference source. }
Training and testing data are generated by spreading the sensing UE terminals in the area uniformly at random, and generating for each pair $(t(n), r(n))$ the CG observation $\tilde c_n := C_{t(n), r(n)} + \epsilon_n$, with $\epsilon_n \sim \mathcal{N}(0, \sigma_c^2)$. The CGs are expressed in dB, and $\sigma_c = 2 dB$.

\begin{figure*}
    \hspace{-2.8cm}
    \includegraphics[width=1.35\textwidth]{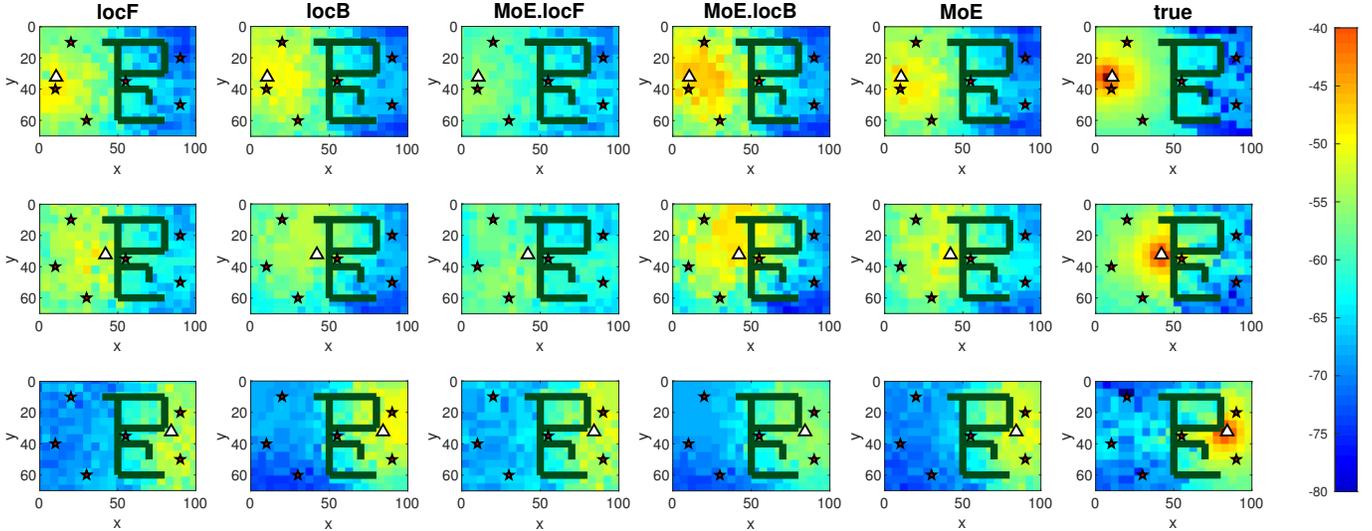}    
    \caption{\small{Colormaps for several slices of the CG map produced by each estimator in experiment 1. Dark lines represent walls, and stars represent positioning pilot signal sources. Each pixel's color indicates the estimated CG in dB between a transmitter located at the triangle and a receiver located at the pixel center. Each column corresponds to a different estimator, where "MoE.locF" ("MoE.locB") denotes the LocFree (LocBased) expert of the mixture; $N_p = 2000$ samples. 
    }}
    \label{fig:storyboard}
\end{figure*}

A first experiment is run to visualize the estimators resulting from the proposed algorithm. Steps 1 and 2 in Algorithm 1 respectively produce a LocF and a LocB CG estimators, which are not part of any mixture. Once the joint optimization of $\bm f_l, \bm f_p, \bm g$ is done (Steps 5-7), not only the final estimator $f(\cdot,\cdot)$ is available, but also $f_l(\cdot,\cdot)$ and $f_p(\cdot,\cdot)$ as a by-product\footnote{\rev{The functions $f$, $f_l$ and $f_p$ are respectively labeled in Fig. 1 as MoE, MoE.locB and MoE.locF.}}, which are different from the estimators trained in Steps 1 and 2. Fig. \ref{fig:storyboard} shows a subset of the estimated CGs for each of these 5 estimators, and also shows the true gains for comparison. \acom{comment further on Figure 1, describing a bit the different channel gain maps we obtain, what features we see, etc.} \rev{It can be observed that, in the first two rows, MoE.locF underestimates the CG in several areas, whereas MoE.locB tends to overestimate them. Interestingly, in the third row one can observe the converse situation. The combination of them shown in the MoE column provides a more balanced CG map.}

\begin{figure}
    \vspace{-1mm}
    \hspace{-5mm}
    \includegraphics[width=0.55\textwidth]{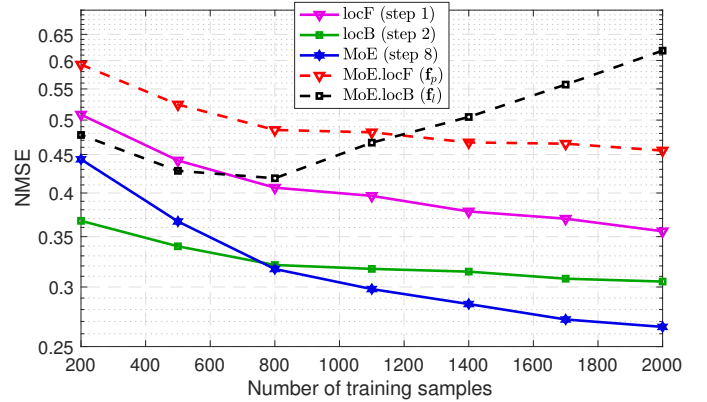}
    \caption{\small{Comparison of test NMSE for several different learning instances with a different number of training samples. Hyperparameters determined by Alg. 1. Results averaged over 30 Monte Carlo realizations.}}
    \label{fig:NMSEvsIterations}
\end{figure}

To analyze the difference in performance between the proposed mixture estimator MoE, and the simple estimators LocB and LocF, a second experiment is run. The goal is to compare the normalized mean square error (NMSE) incurred by each of the aforementioned estimators for different number of training samples, shown in Fig. \ref{fig:NMSEvsIterations}. The NMSE is defined as 
$$\text{NMSE}=\mathbb{E}\{ \vert f(\bm \psi_t, \bm \psi_r)-C_{t,r} \vert ^2 \} / \mathrm{var}\{ C_{t,r}\} $$ 
where the expectation and variance are taken over locations uniformly distributed across the region of interest. The main feature to remark in Fig. \ref{fig:NMSEvsIterations} is that, above a certain number of training samples (800 for this experiment), the MoE estimate (which combines MoE.locB and MoE.locF) achieves a better performance than the (simple) LocF or LocB estimators. This suggests that a training set with too few samples does not carry enough information to successfully learn the three functions involved in MoE. 

\rev{The increase of the NMSE incurred by MoE.locB when the number of samples becomes higher is also remarkable. A possible explanation for this behaviour is that the MoE.locB spatial function becomes more complex (rougher) in an attempt to make the MoE fit the data better. Increasingly complex estimators usually lead to overfitting but, according to Fig. 2, MoE does not overfit. This suggests that the gating function is successfully filtering out the abrupt changes in MoE.locB.   }

\section*{Acknowledgement}
\theAck

\section{Conclusions}
\label{sec:conclusions}
A mixture-of-experts (MoE) model has been proposed to map CG between the locations of any transceiver pair in the area of interest. The location-free (LocF) and location-based (LocB) approaches are combined \rev{using a gating function that incorporates the uncertainty associated with the location estimate. The proposed algorithmic approach learns the MoE.locF and MoE.locB components and the gating function using a block-coordinate minimization approach.} Experiments with simulated data confirm the ability of the proposed approach to perform with lower error than the simple LocF or LocB estimators. \rev{These results motivate future work extending the experimental setup with more realistic and diverse propagation scenarios}.

\if\editmode1 
\printbibliography
\else
\bibliography{\bibfilenames}
  

\fi
\end{document}